# Learning-Assisted Day-Ahead Energy Scheduling for Frequency-Secure Inverter-Dominated Grids with Grid-Forming Battery Energy Storage Systems

Fan Jiang
Department of Electrical and Computer Engineering
*University of Houston*
Houston, TX, USA
fjiang6@uh.edu

Xingpeng Li
Department of Electrical and Computer Engineering
*University of Houston*
Houston, TX, USA
xli83@central.uh.edu

***Abstract*—As grid-forming (GFM) battery energy storage systems (BESS) are increasingly deployed to enhance power system inertial response and frequency stability, incorporating their frequency support capabilities into day-ahead energy scheduling (DAES) is essential for achieving both frequency security and operational efficiency. However, accurately determining frequency metrics in grids with coexisting GFM inverters and synchronous generators requires electromagnetic transient (EMT) simulations, which are computationally prohibitive for direct embedding in grid operational optimization models. To bridge the gap between modeling accuracy and computational efficiency, a learning-assisted DAES (LA-DAES) framework is proposed in this work. By leveraging a surrogate model to represent the frequency support dynamics of GFM BESS, the proposed framework ensures frequency security with a reasonable solve time. Comparative results demonstrate that, relative to analytical frequency-constrained DAES, the proposed LA-DAES framework more accurately captures grid frequency metrics and improves the utilization of GFM BESS.**

## I. Introduction

Global decarbonization demands the transition of power systems toward inverter-dominated grids to accommodate a higher share of renewable energy and inverter-based loads. However, replacing synchronous generators (SGs) with grid-following inverter-based resources reduces system inertia, thereby compromising system frequency stability [1]. The emergence of grid-forming (GFM) inverters offers a promising approach to address this challenge [2]. While extensive literature exists on GFM inverter control strategies for different resources [3]-[5], GFM battery energy storage systems (BESS) currently dominate real-world industrial deployments. Recent deployments and field events have demonstrated the practical potential of GFM BESS for supporting frequency stability in low-inertia grids [6]-[9]. However, to provide effective frequency support following *N*-1 contingency events, GFM BESS must maintain power reserve and sufficient ramping capability [10], which inevitably compromises the economic efficiency of the overall power system. Therefore, to strike a balance between frequency stability and cost effectiveness, the frequency support capabilities and reserve requirements of GFM BESS should be explicitly incorporated into the day-ahead energy scheduling (DAES) framework.

A key challenge in incorporating dynamic frequency response into DAES lies in the simultaneous modeling of cross-timescale features. Specifically, DAES is an optimization framework that characterizes the steady-state behavior of the system at one-hour resolution, whereas frequency response is governed by transient dynamics that unfold within seconds. The fast transient behavior is described by high-dimensional differential-algebraic equations that are highly nonlinear and non-convex, rendering direct integration of frequency metrics such as change of frequency (RoCoF) and frequency nadir (FN) into a scheduling model computationally intractable.

To address this issue, analytical methods have been widely adopted to approximate and linearize the transient frequency response. [11] derived analytical formulations for RoCoF and FN by leveraging the simplified single-machine equivalent system frequency response (SFR) model under the center of inertia framework. Similarly, [12] constrained FN by fitting the classical governor response curve. However, the lumped frequency model used in [11]-[12] fail to capture the spatial distribution of nodal frequencies. To remedy this, [13]-[14] captured the spatial variations of RoCoF via piece-wise linearized local constraints considering system oscillation. Furthermore, [15] discretized and smoothed the differential-algebraic equations, transforming the frequency-constrained scheduling into nonlinear programming problem solvable by commercial solvers. However, these analytical approaches are derived from electromechanical-scale models and electromagnetic transient (EMT) simulations provide high-fidelity characterization of fast GFM control dynamics, including power limiting and inner-loop effects, but are computationally prohibitive for direct embedding

Beyond analytical approaches, deep learning (DL) technologies [16] provides an alternative approach as data-driven surrogates to predict RoCoF and FN accurately. [17] applied DL model to predict FN while considering the inertia contribution of GFM inverters for virtual inertia scheduling. An active sampling DL model was proposed in [18] to predict FN within a DAES framework, and [19]-[21] developed a DL surrogate to constrain both RoCoF and FN. However, these data-driven scheduling methods typically assume that the GFM source possesses infinite power and energy capacity during the transient response, largely neglecting the physical limitations of GFM BESS.

Although the daily energy and power capacity constraints of BESS have been widely incorporated into DAES frameworks [22], the transient energy depletion and ramping limitations during frequency events, which are tightly coupled with GFM control dynamics, remain poorly modeled, causing potentially inaccuracy of frequency response capture. To bridge this gap, this paper proposes a learning-assisted day-ahead energy scheduling (LA-DAES) framework with three key contributions:

- A DL surrogate trained on EMT simulation data is developed to simultaneously predict transient frequency response metrics (RoCoF and FN) and the dynamic reserve requirements of GFM BESS, accounting for the fast control dynamics that are only observable in EMT simulations.
- Transient power reserve and ramping constraints of GFM BESS are explicitly formulated within the LA-DAES framework, establishing a cross-timescale link between second-scale frequency dynamics and hourly energy scheduling.
- The proposed LA-DAES framework is validated against benchmark models, demonstrating that it successfully guarantees frequency stability while ensuring optimal and safe utilization of GFM BESS.

## II. Day-Ahead Energy Scheduling Framework

This section presents the traditional day-ahead energy scheduling (T-DAES) model with BESSs, without considering frequency-related constraints, serving as a benchmark.

In this paper, linear generation cost functions are adopted for SGs. [14]The objective function of the DAES problem is formulated as:

$$min \sum_{g\in\mathcal{G},t\in T}(c_g P_{g,t} + c_g^{\mathrm{NL}} u_{g,t} + c_g^{SU} v_{g,t}) \quad (1)$$

where $\mathcal{G}$ denotes the set of SGs, and $T$ denotes the set of time intervals. $c_g$, $c_g^{\mathrm{NL}}$, $c_g^{SU}$ are the cost coefficient, no-load cost and start-up cost of SG $g$, respectively. $P_{g,t}$ is the active power output of SG $g$ at time interval $t$. $u_{g,t}$ and $v_{g,t}$ denote the commitment status and start-up status of the SG $g$ at time interval $t$, respectively, both of which are binary variables.

The physical constraints of SGs include the active power output limits in (2), the spinning reserve limits in (3)-(4), the ramp-up and ramp-down limits in (5)-(6), the minimum up-time and down-time constraints in (7)-(8), and the logical relationship between the commitment and start-up variables in (9).

$$P_g^{\min} u_{g,t} \le P_{g,t} \le P_g^{\max} u_{g,t} - r_{g,t}, \forall g\in\mathcal{G}, t\in T \quad (2)$$

$$0 \le r_{g,t} \le R_g^{10} u_{g,t}, \forall g\in\mathcal{G}, t\in T \quad (3)$$

$$\sum_{i\in\mathcal{G}, i\neq g} r_{i,t} \ge P_{g,t}, \forall g\in\mathcal{G}, t\in T \quad (4)$$

$$P_{g,t} - P_{g,t-1} \le R_g u_{g,t-1} + R_g^{SU} v_{g,t},\ \forall g\in\mathcal{G}, t\in T \quad (5)$$

$$P_{g,t-1} - P_{g,t} \le R_g u_{g,t} + R_g^{SD}\left(v_{g,t} - u_{g,t} + u_{g,t-1}\right), \forall g\in\mathcal{G}, t\in T \quad (6)$$

$$\sum_{p=t-T_g^U+1}^{t} v_{g,p} \le u_{g,t},\ \forall g\in\mathcal{G}, t\in [T_g^U, nT] \quad (7)$$

$$\sum_{p=t+1}^{t+T_g^D} v_{g,p} \le 1 - u_{g,t},\ \forall g\in\mathcal{G}, t\in [1, nT - T_g^D] \quad (8)$$

$$u_{g,t} - u_{g,t-1} \le v_{g,t},\ \forall g\in\mathcal{G}, t\in T \quad (9)$$

where $P_g^{\min}$ and $P_g^{\max}$ are the minimum and maximum active power limits of the SG $g$, respectively. $r_{g,t}$ denotes the spinning reserve provided by the SG $g$ at time interval $t$. $R_g$ is the normal ramping limit. $R_g^{SU}$ and $R_g^{SD}$ are start-up and shut-down ramping limits of SG $g$, respectively. $R_g^{10}$denotes the ramping capacity of SG $g$ over a 10-minute interval. $T_g^U$ and $T_g^D$ are the minimum up-time and minimum down-time requirements of SG $g$, respectively. $nT$ denotes the total number of time intervals.

It should be noted that the reserve constraint only represents the post-contingency steady-state active power requirement and does not capture transient frequency dynamics.

In addition to the SG constraints, the nodal power balance constraint is imposed as

$$\sum_{k\in\mathcal{K}^-(b)} P_{k,t} + P_{b,t}^{\mathrm{load}} + \sum_{s\in\mathcal{S}(b)} P_{s,t}^{c} = \sum_{g\in\mathcal{G}(b)} P_{g,t} + \sum_{m\in\mathcal{M}(b)} P_{m,t}^{pv} + \sum_{s\in\mathcal{S}(b)} P_{s,t}^{disc} + \sum_{k\in\mathcal{K}^+(b)} P_{k,t}, \forall b,t \quad (10)$$

where $\mathcal{B}$ denotes the set of buses. $\mathcal{K}^-(b)$ and $\mathcal{K}^+(b)$ are sets of lines with active power flowing into and out of bus $b$, respectively. $\mathcal{G}(b)$, $\mathcal{S}(b)$ and $\mathcal{M}(b)$ denote the sets of SGs, PV units, and BESSs connected to bus $b$, respectively. $P_{b,t}^{\mathrm{load}}$ is the load demand at bus $b$ during time interval $t$. $P_{s,t}^{c}$ and $P_{s,t}^{disc}$ are the charging and discharging power of BESS $s$, respectively; $P_{m,t}^{pv}$ denotes the active power output of PV unit $m$.

The DC power flow model in this paper is expressed as (11).

$$P_{k,t} = \frac{\theta_{p,t} - \theta_{q,t}}{x_k}, p,q\in B,\ \forall k\in\mathcal{K}, t\in T \quad (11)$$

where $\mathcal{K}$ is the set of lines. $\theta_{p,t}$ and $\theta_{q,t}$ are the voltage phase angle at line $k$'s from-bus $p$ and to-bus $q$, respectively, and $x_k$ is the reactance of line $k$. A reference bus angle is fixed to eliminate the degree of freedom in the DC power flow model.

The line thermal limits are given by (12).

$$-P_k^{\mathrm{thm}} \le P_{k,t} \le P_k^{\mathrm{thm}},\ \forall k\in\mathcal{K}, t\in T \quad (12)$$

where $P_k^{\mathrm{thm}}$ is the thermal limit of line $k$.

The PV output can be constrained by (13).

$$0 \le P_{m,t}^{pv} \le P_{m,t}^{pv,\max}, \forall m\in\mathcal{M}, t\in T \quad (13)$$

where $P_{m,t}^{pv,\max}$ is the maximum power of PV unit $m$ at time $t$.

The physical constraints of BESSs are also incorporated into T-DAES, including the charging and discharging status constraint in (14), the charging and discharging power limits in (15)-(16), and the state-of-charge (SOC) constraints in (17)-(19).

$$u_{s,t}^{c} + u_{s,t}^{disc} \le 1, \forall s\in\mathcal{S}, t\in T \quad (14)$$

$$u_{s,t}^{c} P_s^{Bmin} \le P_{s,t}^{c} \le u_{s,t}^{c} P_s^{Bmax}, \forall s\in\mathcal{S}, t\in T \quad (15)$$

$$u_{s,t}^{disc} P_s^{Bmin} \le P_{s,t}^{disc} \le u_{s,t}^{disc} P_s^{Bmax}, \forall s\in\mathcal{S}, t\in T \quad (16)$$

$$SOC_s^{min} \le SOC_{s,t} \le SOC_s^{max}, \forall s\in\mathcal{S}, t\in T \quad (17)$$

$$SOC_{s,t} E_s^{max} = SOC_{s,t-1} E_s^{max} + \Delta t\left(P_{s,t}^{c}\eta_s^{c} - \frac{P_{s,t}^{disc}}{\eta_s^{disc}}\right), \forall s,t \quad (18)$$

$$SOC_{s,0} = SOC_{s,nT}, \forall s\in\mathcal{S} \quad (19)$$

where $u_{s,t}^{c}$ and $u_{s,t}^{disc}$ are charge and discharge status variables of BESS $s$. $P_s^{Bmin}$ and $P_s^{Bmax}$ are the maximum charging or discharging power of BESS $s$. $SOC_s^{min}$ and $SOC_s^{max}$ are the minimum and maximum SOC limits of BESS $s$, respectively. $E_s^{max}$ denotes the energy capacity of BESS $s$; $\eta_s^{c}$ and $\eta_s^{disc}$ are the charging and discharging efficiencies, respectively, and $\Delta t$ is the duration of each time interval.

The above BESS constraints describe its conventional energy scheduling behavior. When GFM BESSs and their transient frequency response capabilities are considered, additional constraints are required, which will be introduced in Section III.C.

Since the T-DAES formulation contains binary commitment, start-up, and BESS operating status variables, it is formulated as a mixed-integer linear programming (MILP) problem and can be solved using commercial optimization solvers.

## III. Learning-Assisted Day-Ahead Energy Scheduling

This section proposes a learning-assisted day-ahead energy scheduling (LA-DAES) framework that incorporates transient frequency-security constraints into the traditional DAES formulation. The structure of the DL surrogate, the surrogate training process, and the reformulation used to embed the surrogate into the DAES problem are introduced in this section.

### *A. Structure of the DL surrogate*

The DL surrogate is developed to predict transient frequency-security metrics and the corresponding reserve requirements of GFM BESSs. Specifically, the surrogate predicts the system-level worst-case rate of change of frequency (RoCoF), the worst-case frequency nadir (FN) when considering spatial distribution, the required active power response of each GFM BESS, and the corresponding short-term reserve energy requirement.

Since the surrogate is ultimately embedded into the DAES optimization problem, its complexity directly affects the computational burden of the proposed LA-DAES framework. Therefore, a fully connected neural network with one hidden layer and rectified linear unit (ReLU) activation functions is adopted. This structure provides a balance between prediction capability and computational tractability.

*1) Inputs of the surrogate:* The surrogate aims to predict the post-contingency transient frequency response using pre-contingency operating conditions. Therefore, the input vector should include the variables that significantly affect frequency dynamics, including the load demand at each bus, the active-power output of each SG, the pre-contingency active-power injection of each GFM inverter, and the location of the *N*-1 contingency. The input vector at time interval *t* is expressed as

$$\mathbf{x}_t = \left[\{P_{b,t}^{load}\}_{b\in B}, \{P_{g,t}\}_{g\in G}, \{P_{s,t}^{GFM}\}_{s\in S}, l_t^{con}\right]^\top \quad (20)$$

where $l_t^{con}$ is the encoded location of the considered contingency.

The *N*-1 contingency is defined as the outage of an online SG, and the outage of the SG with the largest capacity is considered as a representative contingency, and its location is represented using a one-hot encoding vector. If the critical contingency changes with commitment decisions, the encoding is updated according to the online SGs at each time interval.

For a GFM BESS coupled with a PV array on the DC side and sharing the same GFM inverter, the pre-contingency active power injection of the GFM inverter is given by

$$P_{s,t}^{GFM} = P_{s,t}^{disc} - P_{s,t}^{c} + P_{s}^{pv}, \forall s \in \mathcal{S}, t \in T \quad (21)$$

The charging and discharging status variables are not explicitly included in the surrogate input because they can be inferred from $P_{s,t}^{disc}$ and $P_{s,t}^{c}$. Similarly, the SG commitment status is implicitly represented by the SG active power output. This design reduces the dimensionality of the input vector while preserving the key information required for frequency-response prediction. The dimension of the input vector depends on the system scale; systems with more buses, generators, and GFM BESSs require higher-dimensional input vectors.

*2) Outputs of the surrogate:* The output vector of the surrogate is defined as

$$\hat{\mathbf{y}}_t = \left[\text{RoCoF}_t^{out}, \text{FN}_t^{out}, \{P_{s,t}^{out}\}_{s\in S}, \{E_{s,t}^{out}\}_{s\in S}\right]^\top \quad (22)$$

where $\text{RoCoF}_t^{out}$ denotes the predicted worst-case RoCoF, $\text{FN}_t^{out}$ is the predict worst-case FN, $P_{s,t}^{out}$ is the required additional maximum active power response of GFM BESS *s*, $E_{s,t}^{out}$ represents to the corresponding reserve energy required during the initial transient response period.

To capture the spatial distribution of frequency dynamics, the worst RoCoF and FN among all buses are selected as the system-level frequency-security labels. The RoCoF measurement window is typically selected between 5 and 10 cycles [23]; in this work, a 10-cycle window (0.167 s for a 60 Hz system) is adopted, and the worst measured value within this window is used as the RoCoF label.

PV units are assumed to operate at their maximum power point unless curtailment is necessary due to grid reliability concerns and do not provide upward transient active-power reserve, following industry practices. Therefore, the transient frequency support of the GFM inverter is mainly provided by the associated BESS. After an SG outage, the BESS active power response is defined with respect to its pre-contingency operating point, allowing both increased discharging and reduced charging to contribute to upward frequency support. The required active power response $P_{s,t}^{out}$ is defined as the difference between the maximum active power output of BESS *s* during the transient response and its pre-contingency operating point. The required reserve energy $E_{s,t}^{out}$ is defined as the additional energy delivered by BESS (s) during the first 5s after the contingency.

### *B. Training process of the DL surrogate*

EMT simulations provide high-fidelity characterization of GFM BESSs dynamics. Therefore, simulation results obtained from a PSCAD/EMTDC model are used to construct the training dataset for the DL surrogate. Each sample maps the pre-contingency operating condition $\mathbf{x}_t$, including the active power set points of SGs, GFM BESSs, PV units, and bus loads, to the output vector $\hat{\mathbf{y}}_t$. For each operating point, the outage of the online SG with the largest capacity is simulated, and the worst-case RoCoF, FN, required active power response $P_{s,t}^{out}$, and reserve energy $E_{s,t}^{out}$ of each GFM BESS are recorded.

To improve data efficiency and ensure that the samples cover the operating region relevant to LA-DAES, T-DAES solutions are used as baseline operating points. For different load and PV scenarios, the corresponding T-DAES dispatch results are passed to the EMT simulation model to generate labeled samples. Independent perturbations are further added to bus-level load and PV profiles to increase the diversity of operating conditions and improve the generalization capability of the surrogate. The resulting dataset is then divided into training, validation, and testing subsets for model development and evaluation.

## C. Incorporation of the DL surrogate into LA-DAES

Embedding the DL surrogate into the DAES model requires two steps. First, the surrogate itself is reformulated as a set of mixed-integer linear constraints. Second, the outputs of the surrogate are constrained to satisfy frequency-security and BESS reserve requirements.

The DL surrogate can be expressed by (23)-(26).

$$a_0 = x_t \tag{23}$$

$$z_m = a_{m-1}W_m + b_m, m \in [1, M-1] \tag{24}$$

$$a_m = \max(z_m, 0), m \in [1, M-2] \tag{25}$$

$$\hat{y}_t = z_{M-1} \tag{26}$$

where $a_0$ denote the input vector. $M$is the total number of layers. $W_m$ and $b_m$ are the weight matrix and bias vector of layer $m$, respectively. $a_m$ denotes the activation values of neurons. $z_m$ is the pre-activated value.

The ReLU function in (25) is nonlinear. To embed it into DAES formulation, it can be exactly reformulated by the BigM method as follows.

$$a_m \le z_m - h_l(1 - B_m) \tag{27}$$

$$a_m \ge z_m \tag{28}$$

$$a_m \le h_u B_m \tag{29}$$

$$a_m \ge 0 \tag{30}$$

where $h_l$ and $h_u$ are valid lower and upper bounds of $z_m$, respectively, and $B_m$ is a binary variable.

After the DL surrogate is reformulated, its outputs are constrained to ensure transient frequency security and sufficient BESS reserve. The RoCoF and FN can be constrained by

$$RoCoF_t^{out} \le R_{lmt}, \forall t \in T \tag{31}$$

$$FN_t^{out} \ge f_{lmt}, \forall t \in T \tag{32}$$

where $R_{lmt}$ denotes the RoCoF security limit, $f_{lmt}$ represents the FN security limit. Note that RoCoF is treated as an absolute value.

The active power response and reserve energy requirement of each GFM BESS are constrained by its available power headroom and SOC margin, respectively:

$$P_{s,t}^{out} \le P_s^{max} - P_{s,t}^{disc} + P_{s,t}^{c}, \forall s \in \mathcal{S}, t \in T \tag{33}$$

$$E_{s,t}^{out} \le SOC_{s,t}E_s^{max} - SOC_s^{min}E_s^{max}, \forall s \in \mathcal{S}, t \in T \tag{34}$$

By integrating (23)-(24) and (26)-(34) into the T-DAES formulation, the proposed LA-DAES model can be formulated as an MILP problem and solved using commercial solvers.

# IV. Case Studies

This section evaluates the performance of the proposed LA-DAES framework on the modified IEEE 9-bus system against two benchmark models: (i) T-DAES mentioned in Section II without frequency constraints; and (ii) linear frequency-constrained (LFC)-DAES which utilizes analytical linear RoCoF and FN constraints from [11].

## A. The modified IEEE 9-bus system

The topology of the modified IEEE 9-bus system is shown in Fig. 1. The system includes four SGs with different capacities and parameters. SG1 and SG2 are located at bus 1, while SG3 and SG4 are located at bus 3. A BESS and a PV plant are connected to bus 2 through an aggregated GFM inverter.

The load and PV profiles are based on 24-hour ERCOT demand and generation data on August 17, 2025. Based on these profiles, the load demand and PV generation are varied within the range of [85%, 115%].

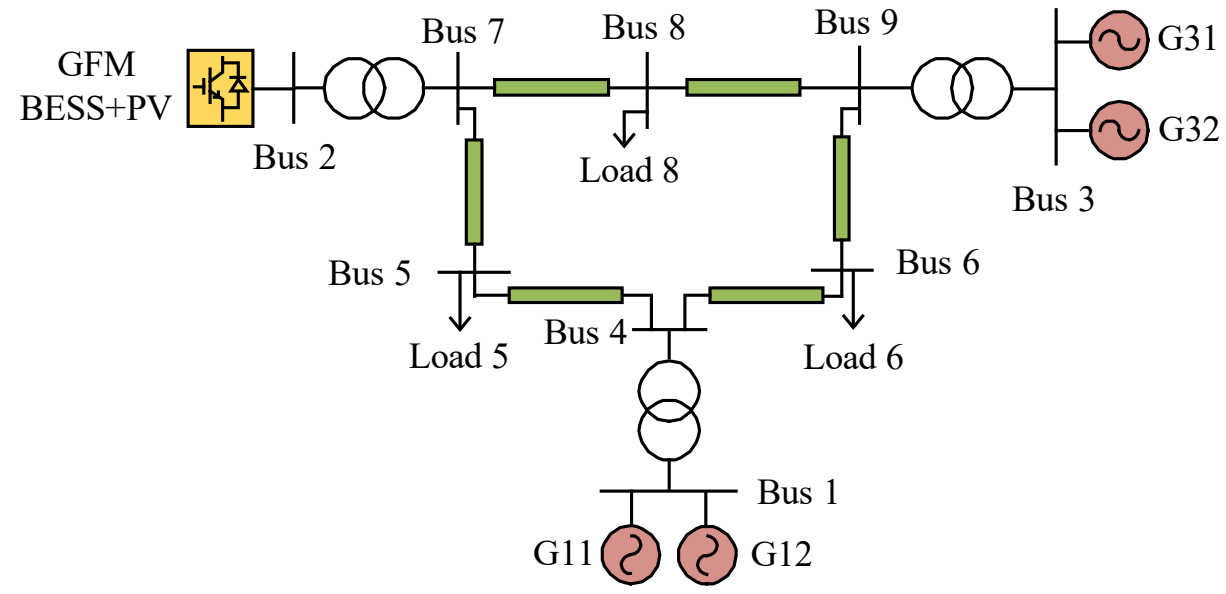

Fig. 1. The modified IEEE 9-bus system.

## B. Development of the DL surrogate

To construct the dataset for the DL surrogate, a PSCAD/EMTDC simulation model is developed. The dataset is then generated following the procedure described in Section III.B. Using this dataset, a DL surrogate with one hidden layer and 32 neurons is trained. Fig. 2 presents the performance evaluation of the proposed surrogate. As shown in Fig. 2(a), both the training and validation losses decrease smoothly and converge to low values. In addition, Fig. 2(b) shows that the normalized predicted values closely match the normalized simulation results that are used as ground-truth labels. These results demonstrate that the proposed DL surrogate achieves satisfactory prediction accuracy.

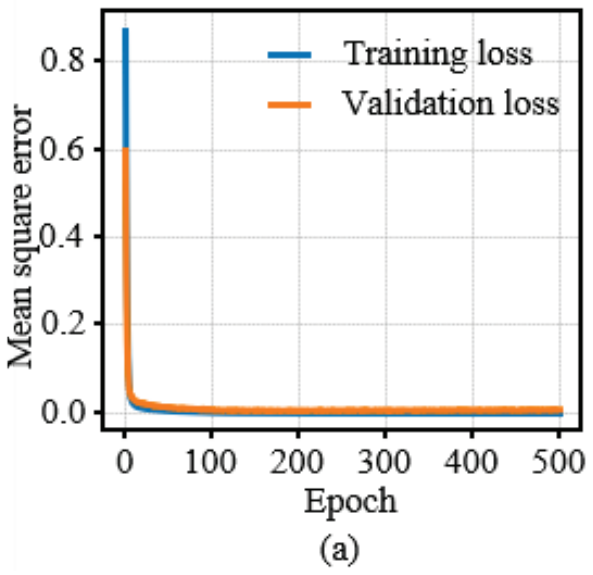

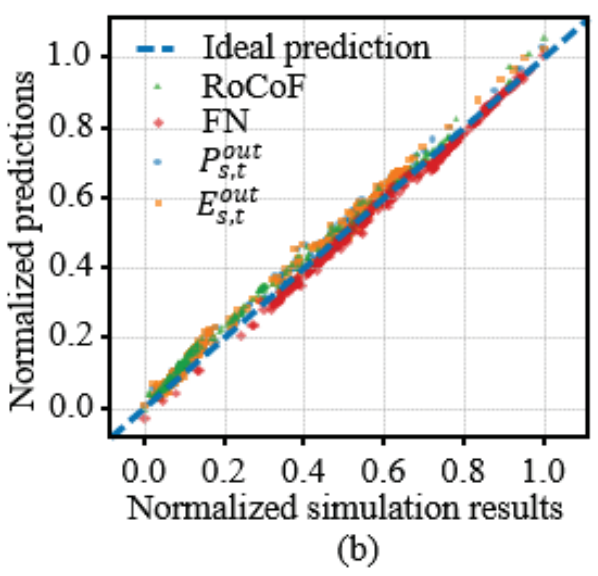

Fig. 2. Performance evaluation of the proposed surrogate: (a) training and validation loss curves, (b) simulation and predicted results comparison.

## C. Results of LA-DAES

After incorporating the proposed DL surrogate into the DAES formulation, the proposed LA-DAES and the two benchmark models are implemented in Pyomo and solved on the same computer with a 12th Gen Intel® Core™ i7-12700 CPU @ 2.1 GHz and 32.0 GB of RAM.

The resulting EMT-based frequency response metrics (i.e. RoCoF and FN) and BESS SOC trajectories of the three models are compared in Fig. 3 and Fig. 4, respectively. As illustrated in Fig. 3, T-DAES violates the safety thresholds for both RoCoF and FN, indicating that neglecting frequency-security constraints may lead to significant security risks. Although both LA-DAES and LFC-DAES maintain FN within the prescribed limit, the proposed LA-DAES achieves better RoCoF performance. In addition, Fig. 4 shows that the BESS SOC under LFC-DAES remains nearly unchanged from hour 1 to hour 20, indicating that the storage flexibility is underutilized. As a result, LFC-DAES relies more heavily on less efficient SG dispatch, leading to a total operating cost of $50,246.29. In contrast, the proposed LA-

DAES achieves a better balance between frequency security and resource utilization, reducing the total cost to $49,943.04.

Therefore, by accurately capturing the frequency-support capability and reserve requirements of GFM BESSs, the proposed LA-DAES model can achieve both frequency stability and cost effectiveness.

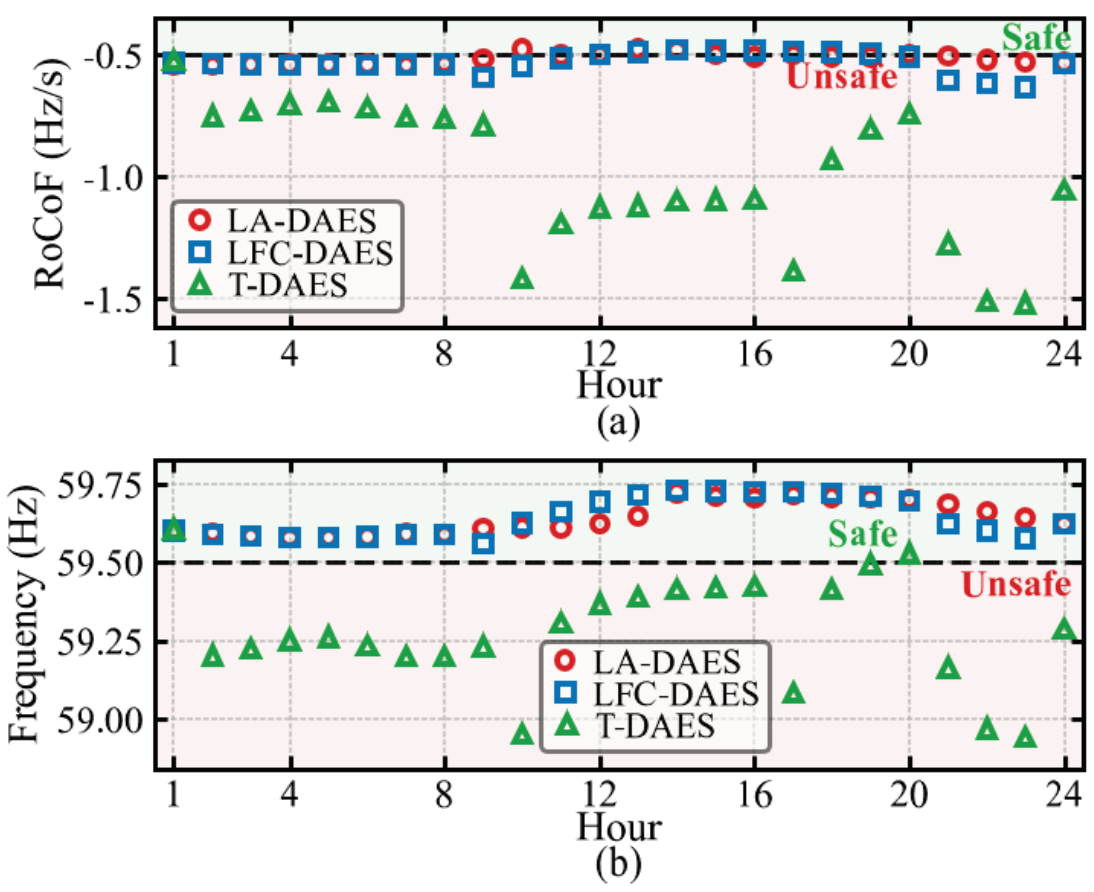

Fig. 3. (a) RoCoF, and (b) FN comparison among proposed LA-DAES, and benchmarks: LFC-DAES and T-DAES.

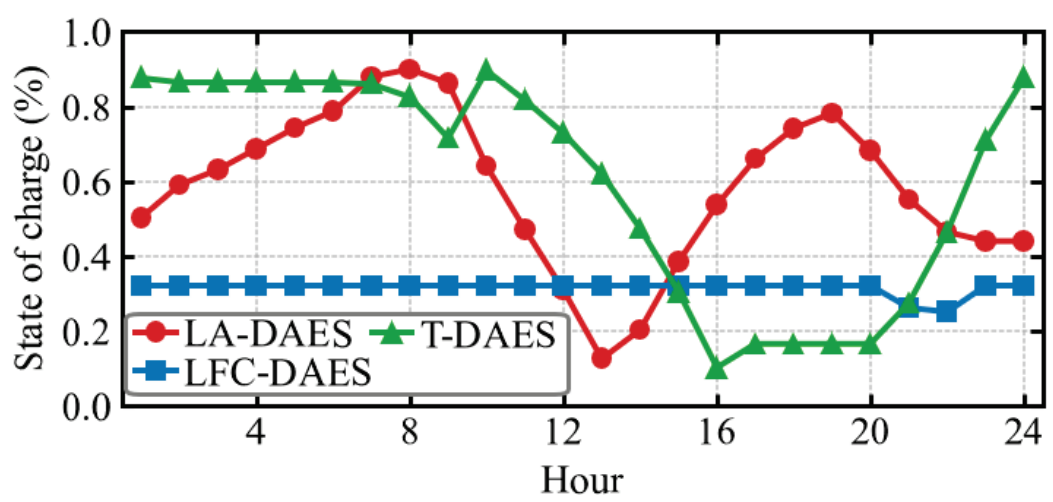

Fig. 4. BESS SOC comparison among the proposed LA-DAES, and benchmarks: LFC-DAES and T-DAES.

## V. Conclusion

This paper proposes an LA-DAES framework that explicitly incorporates the frequency support capability of GFM BESS into power system day-ahead generation scheduling. The results demonstrate that, compared with conventional analytical frequency-constrained approaches, the proposed LA-DAES framework better utilizes GFM BESS flexibility while improving system frequency response performance. By bridging EMT-level accuracy and optimization efficiency, the proposed method can practically maintain frequency security in increasingly inverter-dominated power grids. Future work will focus on further reducing the computational cost when extending the framework to larger systems with multiple GFM BESS units.

## Acknowledgment

This research was in part supported by the National Science Foundation (NSF) under Grant No. 2337598 and the University of Houston Chevron Energy Graduate Fellows Awards.